Scalable Graphene Aptasensors for Drug Quantification

Ramya Vishnubhotla[†,§], Jinglei Ping[†,§], Zhaoli Gao[†], Abigail Lee[†], Olivia Saouaf[†], Amey Vrudhula[ᴜ] and A. T. Charlie Johnson[†*]

[†] Department of Physics and Astronomy, University of Pennsylvania, Philadelphia, PA 19104, USA

[ᴜ] Department of Bioengineering, University of Pennsylvania, Philadelphia, PA 19104, USA

[§] R. Vishnubhotla and J. Ping contributed equally to this work

[*] Corresponding author email: cjohnson@physics.upenn.edu

**Abstract:** Simpler and more rapid approaches for therapeutic drug-level monitoring are highly desirable to enable use at the point-of-care. We have developed an all-electronic approach for detection of the HIV drug tenofovir based on scalable fabrication of arrays of graphene field-effect transistors (GFETs) functionalized with a commercially available DNA aptamer. The shift in the Dirac voltage of the GFETs varied systematically with the concentration of tenofovir in deionized water, with a detection limit less than 1 ng/mL. Tests against a set of negative controls confirmed the specificity of the sensor response. This approach offers the potential for further development into a rapid and convenient point-of-care tool with clinically relevant performance.

Therapeutic drug monitoring (TDM) is crucial for treating patients safely and appropriately as well as for developing new medications. It is particularly important to oversee the consumption of drugs with narrow therapeutic ranges and marked pharmacokinetic variability in target concentrations that are difficult to monitor, and drugs known to cause adverse effects[1] both in individuals and communities. Conventional TDM, however, is based on analytical techniques, such as liquid chromatography and mass spectrometry (LC-MS) that are expensive, time-consuming, and not suitable for clinical use[2]. In this study, we describe the fabrication of nanosensors potentially useful for monitoring the HIV medication tenofovir, with a methodology that leverages the remarkable sensitivity of the two-dimensional material graphene[3], a highly reproducible and robust fabrication method for graphene field effect transistors (GFETs), and an effective, commercially-obtained aptamer with high affinity for tenofovir, a relevant drug metabolite.

Aptamers are oligonucleotide biorecognition elements selected to bind to a particular target[4], for which there are relatively few reports of use with scalable GFETs[5,6,7,8]. It is also possible to integrate aptamer biorecognition layers with metal-oxide-silicon field effect transistors (MOSFETs) using an extended gate geometry[9]. The aptamer used here was obtained commercially (Base Pair Technologies) and has been selected to bind to a metabolite of the HIV market prodrug tenofovir alafenamide. Tenofovir detection is of particular interest as the medication is often used to treat patients affected with HIV by reducing the virus count in the blood of the patient, and therefore decreasing the chance of the development of AIDS. Additionally, hepatitis B virus (HBV) is an accompanying ailment in HIV patients, and tenofovir treatments have shown to reduce the likelihood of HBV forming drug-resistant mutations, making it more suitable for the treatment of HIV than competing drugs[10]. In 2015, Koehn et al reached tenofovir detection limits of 0.5 ng/mL in plasma and cell samples using a method based on liquid chromatography-mass spectrometry (LC-MS)[11]. Such testing is potentially useful for monitoring therapy and to prevent drug accumulation and toxicity in patients with kidney or liver problems. However, despite the fact that this detection limit is much more sensitive than required for TDM of tenofovir, the cost and slow speed of LC-MS make the approach inconvenient for a clinical setting. All-electronic nano-enabled sensors offer a promising pathway towards a low-cost, rapid testing method suitable for use in the clinic or home.

Here we report development of scalable graphene aptasensors for tenofovir based on back-gated GFETs functionalized with a tenofovir aptamer with a limit of detection of approximately 300 pg/mL (~1 nM). We prepared graphene by chemical vapor deposition and fabricated GFETs using a robust and reproducible photolithographic process, with the GFETs showing a high yield (> 90%) and consistent electronic properties[12]. The chemical functionalization procedure provided high surface coverage with the aptamer, as determined using atomic force microscopy (AFM). The aptasensors showed a wide useful range (about a factor of 1000 in concentration) and high selectivity against related drug compounds. Our approach offers the potential for further development into a rapid and convenient point-of-care tool with clinically relevant performance.

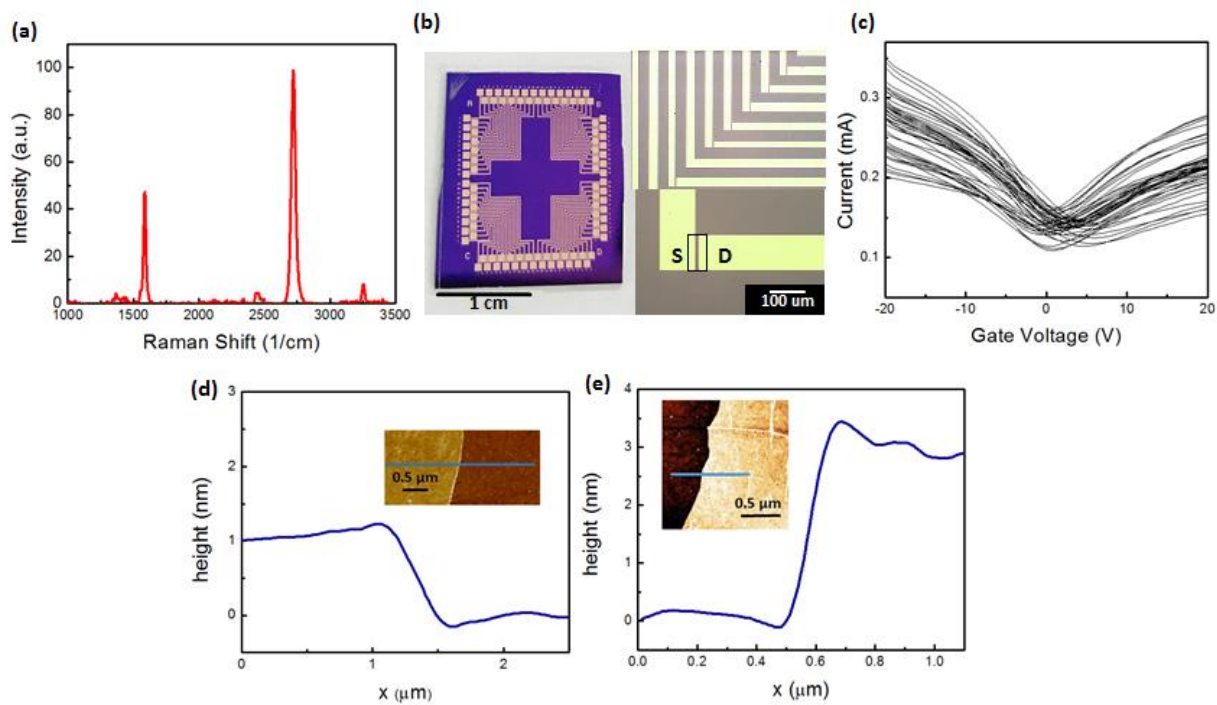

Figure 1. (a) Raman spectrum of chemical-vapor-deposition-grown graphene on copper foil. (b) Three optical images of the sensor array. The left panel is a photograph of an array of 52 graphene field effect transistors (GFETs). The right panel has two optical micrographs at different magnifications. The top micrograph shows a region with vertical source electrodes and horizontal drain electrodes. The lower micrograph is zoomed in to show a single GFET, with a box outlining the graphene channel. (d) Current-gate voltage characteristic of graphene field effect transistors, showing good device uniformity. (e) Atomic Force Microscope (AFM) line scan for annealed graphene on $SiO_2$. The height of the graphene is ~1 nm, as expected for monolayer graphene after transfer onto $SiO_2$. Inset: AFM topographic image with the scan line indicated in blue. (f) AFM line scan of annealed graphene on $SiO_2$ after functionalization with 1-Pyrenebutyric acid N-hydroxysuccinimide ester linker and the tenofovir aptamer. The step height is ~3 nm, consistent with the expected heights for the molecular structure. Inset: AFM topographic image with the scan line shown in blue.

Experiments were based on arrays of 52 devices, with graphene grown by chemical vapor deposition (CVD) on a catalytic copper foil using methane as the carbon feedstock. The monolayer graphene film was transferred onto a pre-patterned array of Cr/Au contacts on an Si/$SiO_2$ wafer (chip size of 2.5 cm x 2 cm) through an electrolysis bubbling method[13]. The quality of the graphene was confirmed *via* Raman spectroscopy (Fig. 1a), showing 2D/G ratio of about 2, as expected for monolayer graphene[14]. GFET channels (10 μm x 100 μm) were defined using photolithography and plasma etching, and the completed GFET arrays (Fig. 1 b,c) were cleaned by annealing in forming gas to minimize contaminants. Additional details of the fabrication are provided in the Methods section.

Current-backgate voltage (I-Vg) measurements showed good device-to-device uniformity across the array (Fig. 1d), and the I-Vg characteristics were analyzed by fitting the data to the form[15]:

$$I^{-1} = \left[e\alpha\mu(V_{bg} - V_D)\right]^{-1} + I_S^{-1} \qquad (1)$$

where $I$ is the measured current, $\mu$ the carrier mobility, $V_{bg}$ the back-gate voltage, $V_D$ the Dirac voltage, α the constant relating gate voltage to carrier number density, and $I_S$ the saturation constant due to short-range scattering[16]. The best fit values for the Dirac voltage and carrier mobility were typically in the range 0-5 V (2.35 ± 1.76 V) with an average mobility of 2,654 ± 115 cm$^2$/V-s.

As-fabricated GFETs were functionalized with a commercial tenofovir aptamer using a well-controlled chemical treatment. First, the GFET array was incubated for ~ 20 hours in a solution of the linker molecule 1-Pyrenebutyric acid N-hydroxysuccinimide ester (P-BASE) at a concentration of 1µM in dimethylformamide (DMF). P-BASE is known to bind with high affinity to graphene *via* π-π stacking[17]. Following the instructions of the manufacturer, the aminated tenofovir aptamer solution (1 µM in phosphate-buffer of pH = 7.6) underwent a heat treatment in order to obtain the desired conformation of the aptamer, and the devices were incubated in this solution for 3 hours following pyrene attachment. Results of the functionalization process were visualized by AFM (Fig. 1 e, f). The height of bare graphene on silicon oxide was ~ 1 nm, while after binding of the linker and aptamer, the height of the structure had increased to ~ 3 nm, consistent with expectations given the molecular structures as well as our earlier report for functionalization of graphene with single-stranded DNA using the same linker molecule[12].

For testing of sensor responses, all 52 aptasensors in a single array were tested against a solution with a known concentration of tenofovir or a related control compound in deionized (DI) water. The solution was pipetted onto the array and left for one hour in order to allow the tenofovir target to bind to the aptamer layer. After incubation, we observed a consistent shift of the Dirac point to more positive gate voltage (Fig. 2a), $\Delta V_D$. The sensor array response was taken to be the average Dirac voltage shift relative to $\Delta V_D^0$, the shift measured upon exposure to deionized water: $\Delta V_D^{REL} = \Delta V_D - \Delta V_D^0$. This relative shift varied systematically with tenofovir concentration (Fig. 2b) and is attributed to an increase in the hole concentration in the GFET due to chemical gating[18] induced by tenofovir binding. Tenofovir contains an amine group and a phosphate group, so it is expected to take on a charge of –e at pH 7.

The Hill-Langmuir model for ligand binding in equilibrium provides an excellent fit to the data for $\Delta V_D^{REL}$ as a function of tenofovir concentration

$$\Delta V_D^{REL} = A \frac{(c/K_a)^n}{1+(c/K_a)^n} + Z \qquad (2)$$

In this equation, *A* represents the maximum response with all binding sites occupied, *c* is the tenofovir concentration, $K_a$ is the tenofovir concentration producing half occupation of a binding site, and *n* is the Hill coefficient. For the data in Fig. 2b, the best-fit parameters are *A*= 9.2 ± 0.2 V, $K_a$ = 3.8 ± 1.5 ng/mL, and *n* = 1.1 ± 0.3, which is consistent with independent binding of the target[19]. Assuming a charge of –e for tenofovir, the shift of ~9 V corresponds to a tenofovir density of 1.1x10³ µm$^{-2}$ when binding is saturated. The GFET tenofovir aptasensors described here have a limit of detection below 1 ng/mL, comparable to that reported for LC-MS, but implemented in a simpler, manufacturable, all-electronic format.

To verify that the sensor response reflected specific binding of tenofovir to the aptamer, tests were conducted against three different HIV drugs as negative controls (lamivudine, abacavir, and emtricitabine), each at a concentration of 200 ng/mL, which for tenofovir would saturate the sensor response. As shown in Fig. 2b, the sensor response to emtricitabine was zero within statistical error, while abacavir and lamivudine gave small but statistically significant responses. This is ascribed to a degree of structural similarity between these compounds and tenofovir that allows for some small binding probability to the aptamer. In a separate control experiment, an array of unfunctionalized graphene FETs was tested against tenofovir at a concentration of 3 µg/mL, a concentration that would saturate the response of the graphene aptasensor. As shown in Fig. 2b, the response of the FET array was zero, within statistical error. Overall the results of these control experiments provide strong evidence that the aptasensor response to tenofovir reflects specific binding to the immobilized aptamer.

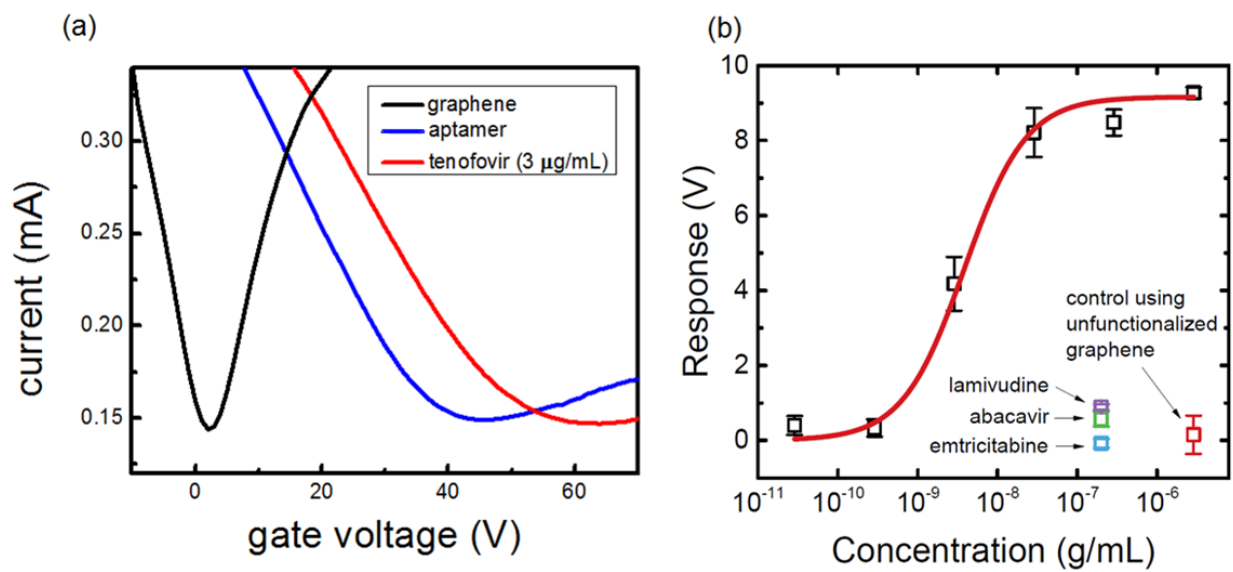

Figure 2. (a) I-Vg curves for an as-fabricated graphene field effect transistor (GFET; black data), the GFET after functionalization with the aptamer (blue data) and after exposure to tenofovir at 3 µg/mL. (b) Relative Dirac voltage shift as a function of tenofovir concentration. The error bars are calculated as the standard error of the mean. The red curve is a fit to the data based upon the Langmuir-Hill model as described in the text. The limit of detection is < 1 ng/mL. Data points associated with negative control experiments are also shown; when no error bar is plotted, the error bar is smaller than the size of the plotted symbol. The near null response for the negative controls provides very strong evidence that the dose-response curve reflects specific binding of the tenofovir target and the aptamer.

**Conclusions**

We have successfully created a scalable approach for fabrication of arrays of GFET-based aptasensors and demonstrated sensitive (~1 nM) and specific detection of the target tenofovir, with a process based on CVD-grown graphene and photolithographic processing, making it suitable for scale-up to industrial production[20]. Our GFET aptasensors have a wide analytical range and sensitivity comparable to LC-MS. Further work is required to optimize the aptasensor performance when applied to real human samples, but their simpler electronic format could make them more suitable for use in a point-of-care setting. For this work, the aptamer was obtained commercially, but we have recently extended the approach to a novel aptamer against azole class antifungal drugs[21], suggesting the ability to incorporate any aptamer into this process.

**Methods**

**Growth of Large-Area graphene by CVD.** CVD graphene was grown in a 4 inch furnace on a copper foil substrate (99.8%, 25 µm, Alfa Aesar) using methane gas as a carbon source and $H_2$ (99.999% pure) as the carrier gas. The foil was placed in a low-pressure, 4" CVD furnace which heated to a temperature of 1020 °C with an $H_2$ flow rate of 80 sccm. After reaching this temperature, the Cu foil was annealed for one hour, with the $H_2$ flow rate constant at 80 sccm. Following the anneal step, methane gas was introduced at 10 sccm, and flowed for 20 minutes. After growth, the furnace was cooled to room temperature before the furnace was vented with $N_2$ gas and the foil was removed.

**Graphene Transfer:** A sacrificial layer of poly methyl methacrylate (PMMA) was spin-coated onto the graphene/copper substrate for structural support, and the graphene was transferred via a hydrolysis bubble transfer method[13] utilizing a 0.05M NaOH solution in deionized water. A potential difference of 20V was applied across the foil as it entered the NaOH bath, with the cathode attached to the foil and the anode in the NaOH solution. As a result, the graphene-PMMA stack lifted off the foil due to the formation of $H_2$ bubbles at the interface of the graphene and copper. After transferring the film to a series of water baths for cleaning, the film was finally transferred onto a 2 x 2.5 cm $SiO_2$/Si chip with a pre-patterned array of Cr/Au (5nm/40nm) metal contact electrodes. The sample was left to dry for ~1 hour, and it was then baked at 150 °C for 2 minutes to further improve adhesion. After this, the PMMA was removed with acetone.

**GFET Fabrication:** A protective layer of polymethylglutarimide (PMGI, Microchem) was spin coated onto the surface of the graphene (4000 rpm, 45 seconds) and baked at 125 °C for 5 minutes. Next, a layer of S1813 photoresist (Microchem) was spin coated onto the sample (5000 rpm, 45 seconds) and baked at 100 °C for 2 minutes. GFET channels were defined using photo-lithography. The excess graphene outside the channels was removed *via* $O_2$ plasma etching (1.25 Torr, 50 W, 30 sec), and the remaining photoresist was removed by soaking the chip in acetone (5 min), 1165 (Microposit, 5 min), and acetone (30 min) before finally being sprayed with isopropyl alcohol (IPA) and dried with compressed $N_2$ gas. The devices were then cleaned of processing residues by annealing in a 1" furnace at ambient pressure under a flow of 250 sscm $H_2$ and 1000 sccm Ar at 225 °C for one hour.

**GFET Functionalization and Testing:** To functionalize the GFET channels, the chip was placed in a solution of 25 mL of dimethylformamide (DMF, Thermo Fisher) and 2 mg of 1-Pyrenebutyric acid N-hydroxysuccinimide ester (P-BASE, Sigma Aldrich), for 20 hours. After this time, the chip was removed, sprayed with DMF and soaked in DMF (2 min), sprayed with IPA and soaked in IPA for 2 min, and finally, sprayed with DI water and soaked in DI water (2 min) before being removed and dried with compressed $N_2$ gas. AFM imaging of samples after this attachment step showed a height of ~2 nm for graphene plus P-BASE (data not shown). To prepare the aptamer solution, 10 µL of a 100 µM aptamer/DI water solution was diluted in 10 mL of phosphate buffer solution ($MgCl_2$, 1mM, pH = 7.4), which was heated from 35 °C to 90 °C, held at 90 °C for 15 minutes, and cooled to room temperature to obtain the necessary configuration of the aptamer. The devices were incubated in this solution for 3 hours. After aptamer attachment (which further increased the AFM height to ~3 nm for graphene, P-BASE and aptamer, as seen in Fig. 1e), the array was thoroughly cleaned with DI water.

The I-Vg curves of the aptamer functionalized GFET array were measured using a bias voltage of 100 mV, while the gate voltage was swept over the range 0 – 90 V, with a step size of 2V and a scan rate of ~ 0.3 V/s. Next, a tenofovir/DI water solution of known concentration was pipetted onto the chip and left to incubate for one hour in a humid environment to prevent evaporation of the solution and allow for specific binding of tenofovir to the aptamer surface. After incubation, the sample was again thoroughly washed with DI water and blown dry. Finally, the I-Vg curves were measured again, and the data was analyzed to determine the Dirac voltage shift due to target binding.

**Ethics Approval:** This work was based on artificial samples so no Ethics Approval was required.

**Acknowledgments**

This work was supported by a grant from the Penn Center for AIDS Research (CFAR), an NIH-funded program (P30 AI 045008). A.L, O.S., and A.V. acknowledge support from the NSF EFRI Research Experience and Mentoring program, through grant EFMA-1542879.

Reference

1.  Andes, D.; Pascual, A.; Marchetti, O., Antifungal therapeutic drug monitoring: established and emerging indications. *Antimicrob. Agents Chemother.* **2009,** *53* (1), 24-34.

2.	Grebe, S. K.; Singh, R. J., LC-MS/MS in the Clinical Laboratory - Where to From Here? *Clin Biochem Rev* **2011,** *32* (1), 5-31.

3.	Chung, C.; Kim, Y.-K.; Shin, D.; Ryoo, S.-R.; Hong, B. H.; Min, D.-H., Biomedical Applications of Graphene and Graphene Oxide. *Accounts of Chemical Research* **2013,** *46* (10), 2211-2224.

4.	Tuerk, C.; Gold, L., Systematic evolution of ligands by exponential enrichment: RNA ligands to bacteriophage T4 DNA polymerase. *Science* **1990,** *249* (4968), 505-10.

5	Kwon, O. S. *et al.* Flexible FET-type VEGF aptasensor based on nitrogen-doped graphene converted from conducting polymer. *ACS Nano* **6**, 1486-1493 (2012).

6.	Ohno, Y.; Maehashi, K.; Matsumoto, K., Label-Free Biosensors Based on Aptamer-Modified Graphene Field-Effect Transistors. *Journal of the American Chemical Society* **2010,** *132* (51), 18012-18013.

7.	Wang, C.; Cui, X.; Li, Y.; Li, H.; Huang, L.; Bi, J.; Luo, J.; Ma, L. Q.; Zhou, W.; Cao, Y.; Wang, B.; Miao, F., A label-free and portable graphene FET aptasensor for children blood lead detection. **2016,** *6*, 21711.

8.	An, J. H., Park, S. J., Kwon, O. S., Bae, J. & Jang, J. High-Performance Flexible Graphene Aptasensor for Mercury Detection in Mussels. *ACS Nano* **7**, (2013).

9.	Aliakbarinodehi, N.; Jolly, P.; Bhalla, N.; Miodek, A.; De Micheli, G.; Estrela, P.; Carrara, S., Aptamer-based field effect biosensor for tenofovir detection. *Scientific Reports* **2017,** *7*, 44409.

10.	Dore, G. J.; Cooper, D. A.; Pozniak, A. L.; DeJesus, E.; Zhong, L.; Miller, M. D.; Lu, B.; Cheng, A. K., Efficacy of tenofovir disoproxil fumarate in antiretroviral therapy-naive and -experienced patients coinfected with HIV-1 and hepatitis B virus. *J Infect Dis* **2004,** *189* (7), 1185-92.

11.	Koehn, J.; Ding, Y.; Freeling, J.; Duan, J.; Ho, R. J., A simple, efficient, and sensitive method for simultaneous detection of anti-HIV drugs atazanavir, ritonavir, and tenofovir by use of liquid chromatography-tandem mass spectrometry. *Antimicrob. Agents Chemother.* **2015,** *59* (11), 6682-8.

12.     Ping, J.; Vishnubhotla, R.; Vrudhula, A.; Johnson, A. T. C., Scalable Production of High-Sensitivity, Label-Free DNA Biosensors Based on Back-Gated Graphene Field Effect Transistors. *ACS Nano* **2016,** *10* (9), 8700-8704.

13.     Gao, L.; Ren, W.; Xu, H.; Jin, L.; Wang, Z.; Ma, T.; Ma, L.-P.; Zhang, Z.; Fu, Q.; Peng, L.-M.; Bao, X.; Cheng, H.-M., Repeated growth and bubbling transfer of graphene with millimetre-size single-crystal grains using platinum. *Nat Commun* **2012,** *3*, 699.

14.     Ferrari, A. C.; Meyer, J. C.; Scardaci, V.; Casiraghi, C.; Lazzeri, M.; Mauri, F.; Piscanec, S.; Jiang, D.; Novoselov, K. S.; Roth, S.; Geim, A. K., Raman spectrum of graphene and graphene layers. *Phys. Rev. Lett.* **2006,** *97*, 187401.

15.     Ping, J.; Xi, J.; Saven, J. G.; Liu, R.; Johnson, A. T. C., Quantifying the effect of ionic screening with protein-decorated graphene transistors. *Biosensors and Bioelectronics*.

16.     Trushin, M.; Schliemann, J., Minimum Electrical and Thermal Conductivity of Graphene: A Quasiclassical Approach. *Physical Review Letters* **2007,** *99* (21), 216602.

17.     Katz, E., Application of bifunctional reagents for immobilization of proteins on a carbon electrode surface: Oriented immobilization of photosynthetic reaction centers. *Journal of Electroanalytical Chemistry* **1994,** *365* (1), 157-164.

18.     Lerner, M.; Resczenski, J.; Amin, A.; Johnson, R.; Goldsmith, J.; Johnson, A., Toward Quantifying the Electrostatic Transduction Mechanism in Carbon Nanotube Molecular Sensors. *J. Am. Chem. Soc.* **2012,** *134* (35), 14318-14321.

19.     Weiss, J. N., The Hill equation revisited: uses and misuses. *The FASEB journal : official publication of the Federation of American Societies for Experimental Biology* **1997,** *11*, 835-841.

20.     Lerner, M. B.; Pan, D.; Gao, Y.; Locascio, L. E.; Lee, K.-Y.; Nokes, J.; Afsahi, S.; Lerner, J. D.; Walker, A.; Collins, P. G.; Oegema, K.; Barron, F.; Goldsmith, B. R., Large scale commercial fabrication of high quality graphene-based assays for biomolecule detection. *Sensors and Actuators B: Chemical* **2017,** *239*, 1261-1267.

21.     Wiedman, g. R.; Zhao, Y.; Mustaev, A.; Ping, J.; Vishnubhotla, R.; Johnson, A. T. C.; Perlin, D. S., An aptamer-biosensor for azole class antifungal drugs. *submitted* **2017**.